\documentclass[11pt, a4paper]{article}
\pdfoutput=1
\usepackage{graphicx}
\usepackage{epsfig}
\usepackage{jcappub}
\usepackage{amsmath}
\usepackage{amssymb}
\usepackage{amsfonts}
\usepackage{epstopdf}

\def \lleq {\lower0.9ex\hbox{ $\buildrel < \over \sim$} ~}
\def \ggeq {\lower0.9ex\hbox{ $\buildrel > \over \sim$} ~}

\def \beq  {\begin{equation}}
\def \eeq  {\end{equation}}
\def \ber  {\begin{eqnarray}}
\def \eer  {\end{eqnarray}}

\newcommand{\be}{\begin{equation}}
\newcommand{\ee}{\end{equation}}
\newcommand{\ba}{\begin{eqnarray}}
\newcommand{\ea}{\end{eqnarray}}
\newcommand{\bea}{\begin{eqnarray*}}
\newcommand{\eea}{\end{eqnarray*}}

\newcommand{\etal}{{\it et al.}}

\usepackage{enumitem}

\title{Bouncing universe in Gauss-Bonnet gravity}

\author{J. K. Singh$ ^1 $,}
\author{Shaily$ ^2 $}
\author{and Kazuharu Bamba$ ^3 $}
\affiliation{$ ^{1,2} $ Department of Mathematics, Netaji Subhas University of Technology, New Delhi-110078, India}
\affiliation{$ ^3 $ Division of Human Support System, Faculty of Symbiotic Systems Science, Fukushima University}
\emailAdd{jksingh@nsut.ac.in}
\emailAdd{shaily.ma19@nsut.ac.in}
\emailAdd{bamba@sss.fukushima-u.ac.jp}

\abstract{ In this paper, a bouncing cosmological scenario is studied in the background of a flat FLRW model with a specific parametrized hyperbolic form of scale factor $ a $ in terms of $ t $, where $ \lambda $ is taken as the model parameter.  This model is discussed in $ f(R,G) $ formalism having structured as $ f(R,G)=R+F(G) $, where $ R $ is Ricci scalar and $ G $ is the Gauss-Bonnet invariant. The proposed functional form of the Hubble parameter is considered in such a way that it satisfies the bouncing criteria of the model, which is free from initial singularity. The physical consequences of the model are discussed. In this model, one can see that the EoS parameter crosses the quintom line $ \omega=-1 $ in the neighbourhood of bouncing point $ t\approx0 $, which is a very strong criterion for a successful bouncing cosmological model. Finally, we find that all the essential features of bouncing model are satisfied successfully.}

\date{\today}

\begin{document}

\maketitle

\section{Introduction}

\qquad Bouncing cosmology has been the subject of interest for many physicists for a long time since we know of the big bang. There are basically some possibilities and ideas that astronomers have about the future of the Universe. There are many competing theories, which predict not only the universal expansion but also its early description and ultimate fate of the Universe. The first idea among these is that there might be so much matter in the Universe that despite the expansion of gravity, would bring everything back into a big crunch\footnote{A cosmic scenario in which everything will be pulled towards each other at some stage and expansion would go in the other direction, and we would have the precise opposite of big bang.}. The second scenario is about the recession of the galaxies from each other and space-time itself expands forever. The last scenario deals with the oscillatory universe, which describes the repeated cycle of the big bang and big crunch and is known as the big bounce, which we shall study in this work.

The big bounce theory is a hypothetical cosmological theory for the origin of the known universe. This theory states that our Universe may experience alternating periods of expansion and contraction moving from one state to another without collapsing upon itself \cite{deHaro:2012xj, Moriconi:2016egx, Wang:2003yr, Ijjas:2016tpn, Battefeld:2014uga}. Many physicists do not support the big bang theory because of the assumption that the Universe was born out of broken physics. Instead, it is the prevailing thought that the current Universe may have formed from an older collapsing Universe or it might always be going through periods of expansion and contraction without collapsing completely. Presently, it is believed that we are in the expansion period and therefore galaxies are getting further away from each other. This theory states that there is nothing like singularity.  

A bouncing cosmological model describes the initial singular state of the Universe as an interesting question raises in modern theoretical cosmology that the universe really starts either with the big bang or with big bounce because it may also be possible that the universe oscillates in a bouncing-like approach. Bouncing cosmology contradicts that an initial singularity exists. In the context of bouncing scenario, the initial singular state of the Universe has been discussed by many authors \cite{Bamba:2014mya,Odintsov:2015zza}. The cosmologists have built their models on the idea that the early Universe at conformal symmetry is governed by the rules of quantum mechanics\footnote{Quantum mechanics are the laws of physics that govern matter at the scale smaller than atoms.}. Quantum mechanics save electrons from falling in and destroying atoms so that it could save the early Universe from such violent beginnings and endings as the big bang and big crunch. Therefore, this theory is considered the solution to the singularity problem in the standard big bang model because quantum mechanics would not allow the small particles to compress into the infinitely dense point with no mass. Bouncing theory can also be undertaken as the consequence of applying `Loop quantum gravity technique' LQG\footnote{LQG is the unified theory that explains the quantum mechanics and the Einstein GR under one umbrella.} to Big Bang Theory \cite{Mielczarek:2008zv, Barrau:2009fz}.

In standard cosmology, some necessary conditions for having a successful bouncing dark energy model are given as follows \cite{Ijjas:2016tpn}:
 
\begin{itemize}[noitemsep]
\item[-] In 4-dimensional FLRW space-time cosmology, the null energy condition (NEC) which is equivalent to $ \dot{H} = -4\pi G \rho (1+\omega) > 0 $ violates in the neighbourhood of a bouncing point.

\item[-] In contracting phase of the Universe, the scale factor $ a(t) $ decreases with respect to $ t $ \textit{i.e.} $ \dot{a}(t)<0 $ and Hubble parameter $ H<0 $. In expanding phase of the Universe, the scale factor $ a(t) $ increases with respect to $ t $ \textit{i.e.} $ \dot{a}(t)>0 $ and Hubble parameter $ H>0 $. Also, $ \dot{a}(t)=0 $ and Hubble parameter $ H=0 $ at bouncing Point.

\item[-] The effective equation of the state (EoS) parameter $ \omega $ crosses the quintom line $ \omega=-1 $. This is an essential critera for the quintom model \cite{Singh:2018xjv, Feng:2004ad}.  
\end{itemize}

While physicists have discussed the idea of a big bounce since long back but this theory has been inhibited by the inability to explain several new cosmological questions due to the inadequacy of many pieces of evidence.
\begin{enumerate}
\item [(i)] If subatomic particles can not compress into a singularity, then what are neutron stars?
\item[(ii)] If the big bang was not the start of everything then what was the start of everything?
\item [(iii)] If singularity does not exist, then what is a black hole?
\end{enumerate}
 
The bouncing behaviour of universe has been discussed by several authors in various alternative theories of gravities like $ f(R) $, $ f(R,T) $,  $ f(G) $, $ f(R,G) $, and $ f(Q,T) $  gravities etc. \cite{Odintsov:2018nch,Odintsov:2015ynk,Elizalde:2020zcb,Nojiri:2022xdo,Odintsov:2020zct,Odintsov:2021yva}. Singh \textit{et al.} \cite{Singh:2018xjv} has discussed bouncing universe on $ f(R,T) $ gravity. B. Mishra discussed the bouncing cosmology in $ f(R,T) $ and $ f(Q,T) $ gravity \cite{Agrawal:2021gdq, Agrawal:2021rur}. Easson \etal ~discussed on G-Bounce cosmology in general relativity (GR) \cite{Easson:2011zy}. Dobre \etal ~ \cite{Dobre:2017pnt} studied the perticular case of bouncing cosmology which was earlier discussed by Ijjas and Steinhardt \cite{Ijjas:2016tpn}. Barca \etal ~ \cite{Barca:2021qdn} reviewed the nature of the Bounce in LQC and PQM.

Many authors have worked on $ f(R,G) $ theory and discussed the energy conditions, future finite time singularities and other cosmological implications. To understand the structure of the Universe and other cosmological behavior, some specific classes of theories are discussed like $ f(R)$  gravity, Hořava–Lifshitz $ f(R) $ gravity, Gauss-Bonnet theory, scalar-tensor theory, non-minimally coupled models and theories,  which contains the higher-derivative gravitational invariants \cite{Nojiri:2010wj,Nojiri:2017ncd,Sotiriou:2008rp,DeFelice:2010aj,delaCruz-Dombriz:2018nvt,Sebastiani:2016ras,Capozziello:2011et, Bamba:2012cp}. In the literature, various dark energy models are discussed to explain the different dynamical features of the Universe. Here, we explore the bouncing cosmological model in $ f(R,G) $ gravity, where $ G $ is the Gauss-Bonnet invariant, defined as $ G=R^2 - 4 R_{ij} R^{ij} + R_{ijkl} R^{ijkl} $, where $ R_{ij} $ is the Ricci tensor and $ R_{ijkl} $ is the Riemann tensor and these are the higher-order corrections on the curvature tensor added to the Einstein’s gravitational action. Using this evolution of the Universe can be explained very clearly as its contribution is more feasible by some string models \cite{Navo:2020eqt}.

In this paper, we study a bouncing cosmological model in $ f(R,G) $ theory of gravity. Sect. II is started with the calculation to find the function $ f(G) $ using reconstructing technique developed in \cite{Odintsov:2018nch}. For this technique,  scalar factor $ a(t) $ is assumed according to the necessary condition of standard cosmology. The gravitational action for $ f(R,G) $ is used to derive the Einstein field equations in Sect. III. Further, the cosmological parameters have been calculated and examined all the necessary conditions to accomplish a bouncing model of the Universe (Sect. IV). Finally, we conclude our results in sect. V. In the last section final result is discussed and summarize this bouncing model.  

\section{Formulation of $ f(R,G) $ gravity using reconstructing technique with $ e $-folding number approach}
\qquad In this section, we introduce the general formalism of $ f(R,G) $ theory of gravity. The gravitational action of vacuum $ f(R,G) $ gravity is defined as

\begin{equation} \label{1}
S=\frac{1}{2 \kappa}\int f(R,G)\sqrt{-g}  d^{4}x,
\end{equation} 

where $ \kappa=8\pi G $ is a constant. In $ f(R,G) $ gravity, function $ R $ is the Ricci scalar and $ G $ is defined as the Gauss-Bonnet invariant as
\begin{equation} \label{2}
G=R^2 - 4 R_{ij} R^{ij} + R_{ijkl} R^{ijkl}, 
\end{equation}

where $ R_{ij} $ is the Ricci tensor and $ R_{ijkl} $ is the Riemann tensor. Now to find the Einstein Field Equations we do the variation in gravitational action with respect to metric tensor $ g_{\mu\nu} $ and find the gravitational equations as

\begin{eqnarray}\label{3}
G_{ij}=\frac{1}{f_R}[\nabla_{\mu} \nabla_{\nu} f_R - g_{\mu \nu} \Box f_R +2R \nabla_{\mu} \nabla_{\nu} f_G - 2g_{\mu\nu} R \Box f_G - 4 R_{\mu}^k \nabla_k \nabla_{\nu} f_G \notag \\ 
- 4 R_{\nu}^k \nabla_k \nabla_{\mu} f_G +  4R_{\mu \nu} \Box f_G + 4 g_{\mu \nu} R_{ij}\nabla_i \nabla_j f_G + 4 R_{\mu i j \nu} \nabla_i \nabla_j f_G \notag \\
- \frac{1}{2} g_{\mu \nu} (R f_R + G f_G - f(R,G)),
\end{eqnarray}

\begin{equation}\label{4}
3\Box f_R + R f_R - 2f(R,G) + R[ \Box f_G + 2G R f_G]=0.
\end{equation}

The gravitational field equations for a flat Friedmann-Lemaitre-Robertson-Walker (FLRW) space-time
\begin{equation}\label{5}
ds^2=-dt^2+a^2 (t)(dx^2+dy^2+dz^2),
\end{equation}
can be written as \cite{Odintsov:2018nch}
\begin{equation}\label{6}
2 \dot{H} f_R + 8H \dot{H} \dot{f_G}=H \dot{f_R}-\ddot{f_R}+4H^3\dot{f_G}-4H^2\ddot{f_G},
\end{equation}
\begin{equation}\label{7}
6H^2 f_R+24H^3\dot{f_G} = f_R R - f(R,G)-6H \dot{f_R} + G f_G.
\end{equation}

Here an overhead dot denotes the derivative with respect to $ t $, $ f_G=\frac{\partial f}{\partial G} $ and $ f_R=\frac{\partial f}{\partial R} $. For FLRW metric, the Ricci tensor $ (R) $ and Gauss Bonnet invariant $ (G) $ are given by
\begin{equation} \label{8}
R=6(2 H^2+ \dot{H}),
\end{equation}
\begin{equation}\label{9}
G=24 H^2(H^2+ \dot{H}).
\end{equation}

In the literature, it has been seen that the analytical solution of the gravitational equations (\ref {6}) and (\ref {7}) are very complicated. So in order to discuss the evolution of universe, we use the reconstructing technique to find the exact solution. This technique has already been used in \cite{Nojiri:2005pu,Bamba:2014daa,Bamba:2014wda} for $ f(R) $ gravity. Our primary aim is to find the value of the function $ f(R,G) $ by assuming the appropriate value of scale factor $ a(t) $. Also, to avoid the complexity in solving the equation we restrict our model for the $ f(R,G) $ gravity as $ f(R,G)=R+F(G) $.\\

In this technique, the Ricci tensor $ R $ and Gauss Bonnet invariant $ G $ can be expressed in the form of e-folding number, $ N=ln(a/a_0) $ or $ dN=H dt $ defined as
\begin{equation} \label{10}
R(N) = 6(2H^2(N)+H(N)H'(N)),
\end{equation}  
\begin{equation} \label{11}
G(N) = 24H^2(N)(H^2(N)+H(N)H'(N)),
\end{equation}  
where $ H'(N)=\frac{\partial H}{\partial N} $. Now $ H $ is a function of e-folding number $ N $. So to find $ f(R,G) $ and to make our calculation easier we choose a specific functional form of Hubble parameter $ H $ using reconstruction technique as \cite{Bamba:2014mya}
\begin{equation} \label{12}
H^2(N)= P(N).
\end{equation}     
Thus, the Eqs. (\ref {10}) and (\ref {11}) can be written in terms of $ P(N) $ as follows:     
\begin{equation}\label{13}
R=12P(N)+3P'(N),
\end{equation} 
\begin{equation}\label{14}
G=24P^2(N)+12P(N)P'(N).       
\end{equation}
Using equation (\ref {12}), the Eq. (\ref {7}) can be exprssed as \cite{Navo:2020eqt}
\begin{equation}\label{15}
P(N)(6f_R + 24P(N) f_{GG} G'(N))-f_R R + f(R,G)+6 P(N) f_{RR} R'(N) - G(N) f_G = 0.
\end{equation}\\
Now since $ f(G,R)=R + F(G) $, therefore $ f_{RR} = 0 $ and $ f_R = 1 $ and the (\ref {15}) can be written as
\begin{equation}\label{16}
P(N)(6 + 24P(N)F_{GG} G'(N)) + F(G) - G(N)F_G = 0.
\end{equation}

In order to construct the function $ F(G) $ from the second order differential Eq. (\ref {16}), let us assume the cosmic scale factor $ a(t)= cosh (\lambda t $), where $ \lambda $ is an arbitrary constant. The corresponding Hubble parameter $ H $ can be calculated as 
\begin{equation}\label{17}
H = \frac{\dot{a}}{a}= \lambda tanh (\lambda t),
\end{equation}
using e-folding number relation, we get 
\begin{equation}\label{18}
H^2(N)=P(N)=\lambda^2(1-e^{-2N}).
\end{equation}   
To know more about the late time universe, one can take $ N \to \infty $ and the corresponding values of $ N(G) $ and $ P(G) $ can be calculated. Using Eq. (\ref {16}), we can get the differential equation as 

\begin{equation}\label{19}
-GF_{GG} + F(G) +\lambda^2 G = 0,
\end{equation}
which gives
\begin{equation}\label{20}
F(G)=aG + \lambda^2 G lnG,
\end{equation}
where $ a $ is the integrating constant. Also the reconstructed value of $ f(R,G) $ gravity can be written as
\begin{equation}\label{21}
f(R,G)=R+aG+\lambda^2 G lnG,
\end{equation}
Now we proceed to discuss the cosmological parameters using this $ f(R,G) $ gravity function.

\section{Einstein Field Equations in $ f(R,G) $ gravity}

\qquad In $ f(R,G) $ gravity, we are trying to explore the cosmological behaviour containing $ S_m $ as the action on the matter sector. In particular, one can drive gravitational modifications using an arbitrary function $ F(G) $  \cite{Odintsov:2015uca,Nojiri:2005vv,Nojiri:2005jg}. The total gravitational action for $ f(R,G) $ gravity can be expressed as
\begin{equation}\label{22}
S=\frac{1}{2 \kappa}\int [R+F(G)]\sqrt{-g}  d^{4}x + S_m,
\end{equation} 
where $ R $ is Ricci scalar and $ G $ is the Gauss-Bonnet invariant defined in (\ref {2}). The Einstein field equations (EFE) for the metric (\ref {5}) yields \cite{Odintsov:2015uca}

\begin{multline}\label{23}
R_{ij}-\frac{1}{2} g_{ij}F(G)+(2RR_{ij}-4R_{ik}R^k_j+2R^{kl\tau}_i R_{jkl\tau}-4g^{\alpha k} g^{\beta l}R_{i\alpha j \beta}R_{kl})F'(G)\\
 +4[\nabla_k \nabla_j F'(G)]R^k_i-4g_{ij}[\nabla_k \nabla_l F'(G)]R^{kl}+4[\nabla_k \nabla_l F'(G)]g^{\alpha k} g^{\beta l}R_{i\alpha j \beta}\\
 -2[\nabla_i \nabla_j F'(G)]R+2g_{ij}[\Box F'(G)]R-4[\Box F'(G)]R_{ij}+4[\nabla_i \nabla_k F'(G)]R^k_j=\kappa T^m_{ij},
\end{multline}
where $ T^m_{ij} $ is the energy momentum tensor arises from the matter sector $ S_m $. For a flat FLRW metric (\ref {5}), the equation (\ref {23}) generates the field equations as
\begin{equation}\label{24}
6H^2+F(G)-GF'(G)+24H^3 \dot{G} F''(G)=2\kappa \rho,
\end{equation}
\begin{multline}\label{25}
4\dot{H} + 6H^2+F(G)-G F'(G)+16 H\dot{G}(\dot{H}+H^2)F''(G)+8H^2\ddot{G} F''(G)\\
+8H^2\dot{G}^2F'''(G)=-2\kappa p,
\end{multline}
where $ \rho $ is the energy density and $ p $ the isotropic pressure for matter. And the value of $ R $ and $ G $ are defined in equation (\ref {8}) and (\ref {9}) respectively.

In the next section, we study the bouncing conditions of our model using the parametrization of cosmic scale factor $ a(t) $ and the Hubble Parameter $ H $ which has been evaluated in Eq. (\ref {17}).   
 
\section{ Dynamical status of bouncing evolution} 

\qquad In the space-time cosmology, the model shows the expanding Universe when $ \dot{a}>0 $ whereas it shows contracting Universe when $ \dot{a}<0 $. Also in the standard Big Bang cosmology, it is assumed that the contraction and expansion are the mirror image about the bouncing point in the absence of entropy production \cite {Brandenberger:2012zb}. To analyse the successful bouncing model we have some necessary conditions in standard cosmology \cite {Cai:2007qw}. Now in order to examine our model in more detail let us explore the the behaviour of various cosmological parameters like   scale factor, Hubble parameter, energy density, isotropic pressure, EoS parameter, and energy conditions \textit {etc.} analytically and graphically. 
 
\subsection{ Hubble Parameter with parametrization of scale factor} 

\qquad In the literature, it is very clear that the scale factor $ a(t) $ and the Hubble parameter $ H $ plays a dominant role to explain the cosmological models. When we frame a bouncing cosmological model, all the necessary conditions should be satisfied, which have been discussed previously. In this paper, we have parametrized the scale factor $ a(t) $ as a function of cosmic time $ t $. To avoid the complexity of the calculations, the Hubble parameter $ H $ in Eq. (\ref {17}) can be taken as:
\begin{equation}\label{26}
H(t)= \lambda  \left(\lambda  t-\frac{\lambda ^3 t^3}{3}\right).
\end{equation}
\begin{figure}
\begin{center}
$%
\begin{array}{c@{\hspace{.1in}}cc}
\includegraphics[width=3.0 in, height=2.2 in]{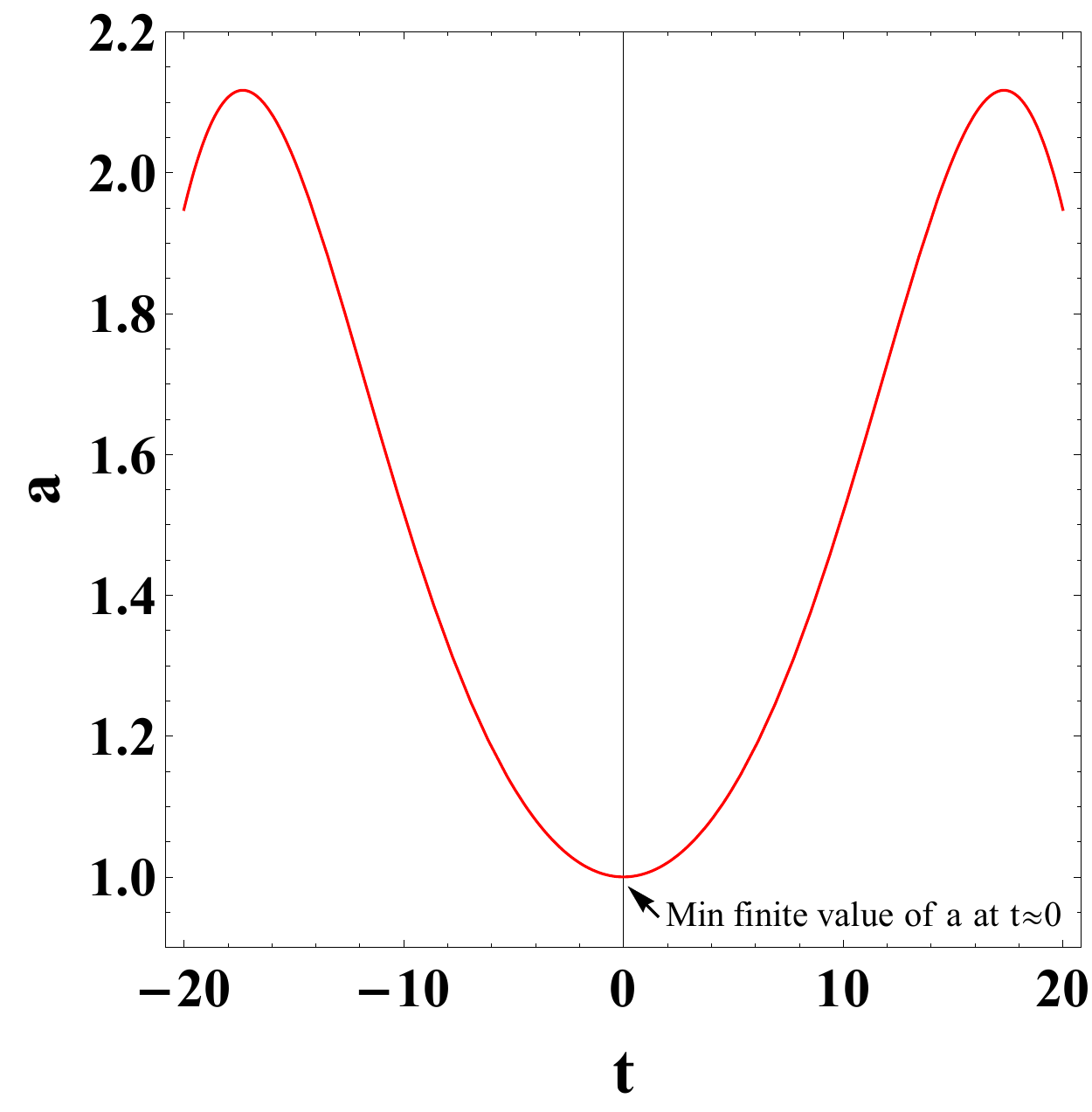} ~~~~~
\includegraphics[width=3.0 in, height=2.2 in]{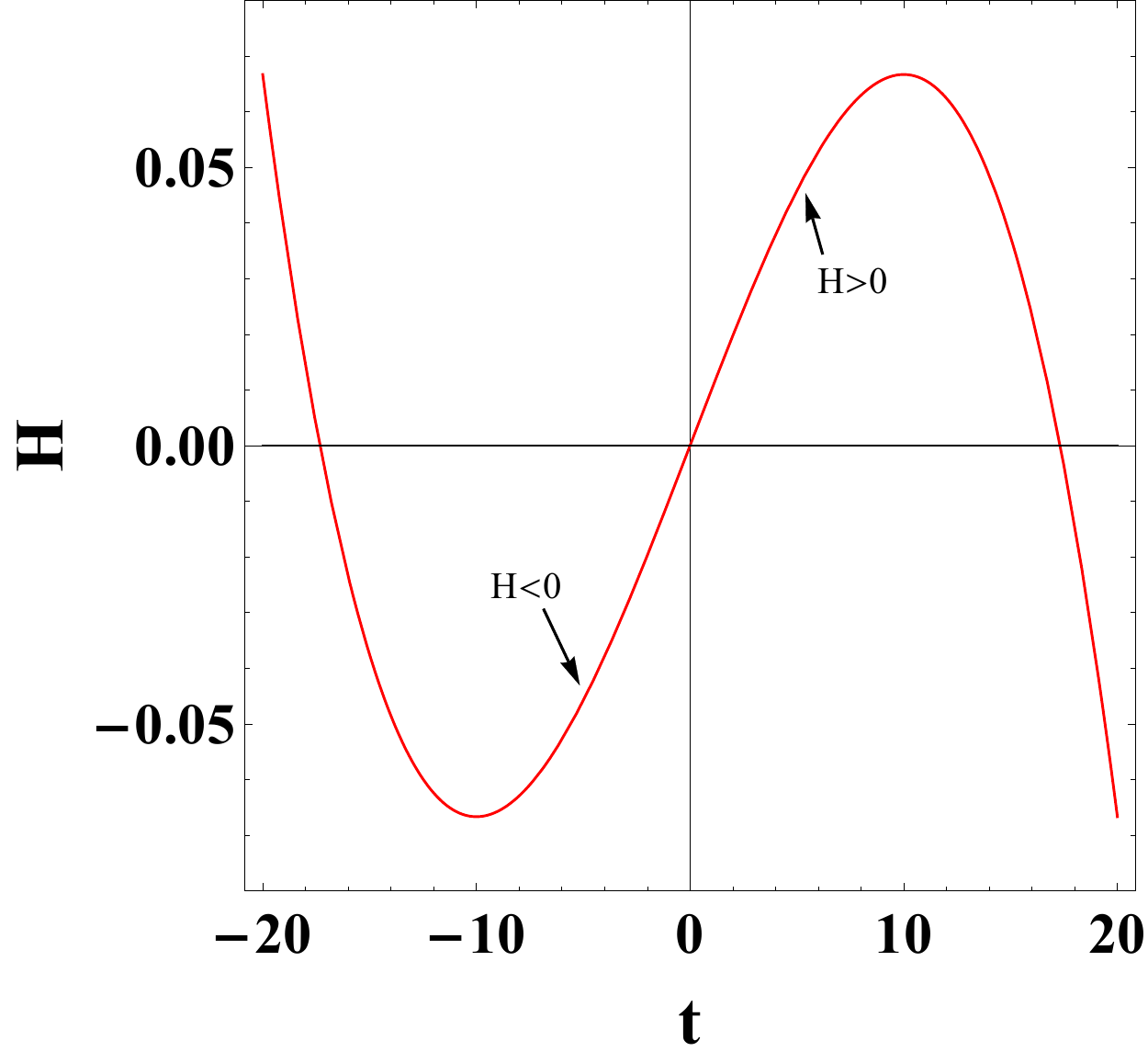} \\
\mbox (a) & \mbox (b) \\
\includegraphics[width=3.0 in, height=2.2 in]{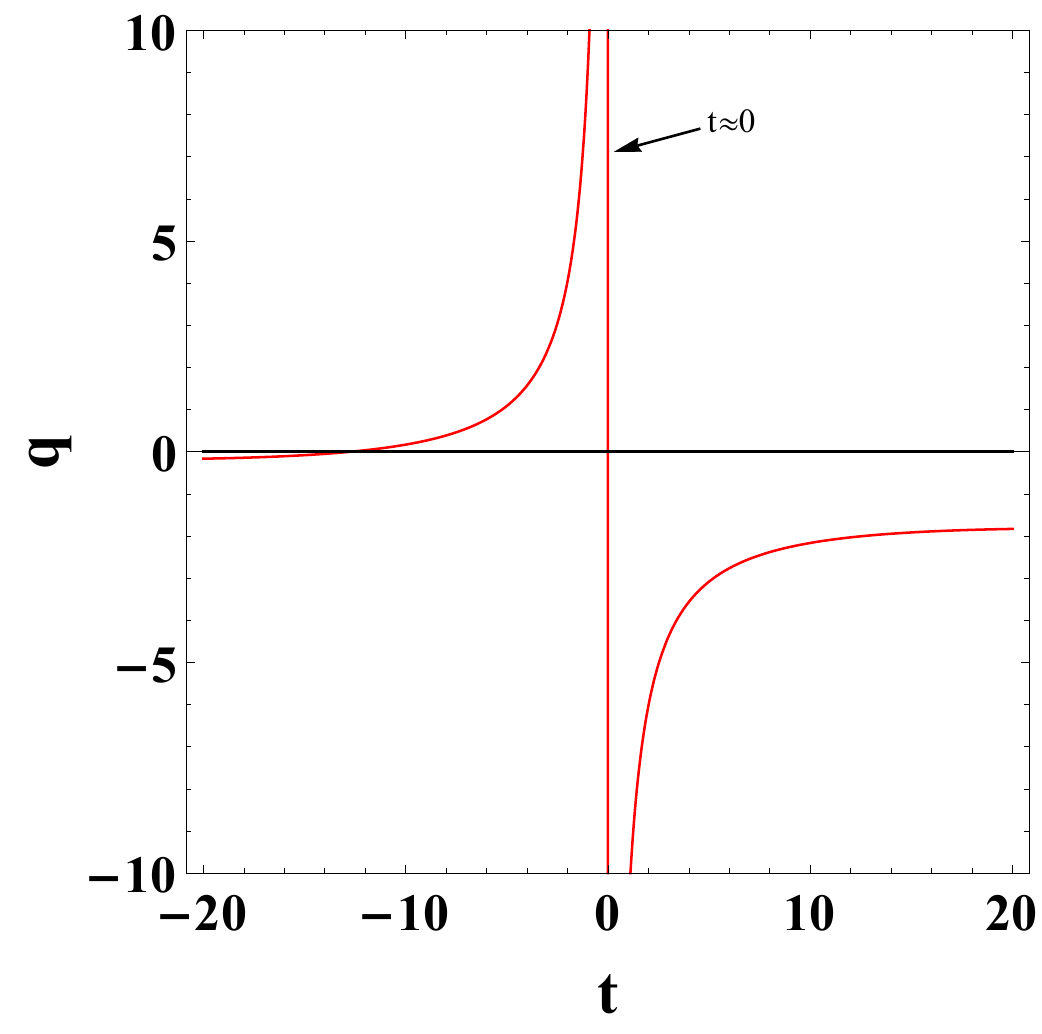} \\ 
\mbox (c)%
\end{array}%
$%
\end{center}
\caption{\scriptsize The graphical behaviours of $ a $, $ H $ and $ q $ \textit{w.r.t.} $ t $.}
\end{figure}

In a cosmic bouncing scenario, the contraction and expansion of universe can be explained with the help of Hubble parameter. The Hubble parameter plot in Fig. 1(b) clearly indicates whether $ H<0 $, $ H>0 $ or the bouncing condition  $ H=0 $. Fig. 1 depicts the curve of scale factor, Hubble parameter and deceleration parameter $ q $ with respect to $ t $ for very small values $ \lambda=0.1 $. The scale factor $ a(t)= \lambda  \left(\frac{\lambda  t^2}{2}-\frac{\lambda ^3 t^4}{12}\right) $ is derived from Eq. (\ref {26}). During the contracting universe, the scale factor $ a(t) $ is monotonically decreasing \textit{ i.e.} $ \dot{a(t)} < 0 $, whereas the scale factor $ a(t) $ indicates increasing pattern, \textit{i.e.} $ \dot{a(t)}> 0 $ during the expanding phase of the Universe. The scale factor of the Universe reaches to a non-zero minimum value $ a = 1 $ at the transfer position $ t=0 $ for $ \lambda=0.1 $ (see Fig. 1(a)). Thus, we see that the volume of the model is decreasing before the bounce and starts increasing after the bounce.
 
Fig. 1(b) depicts that the Universe contracts in the interval $ -17.4 \leqslant t \leqslant 0 $ and then expansion starts at $ t=0 $. The bouncing point of the model is determined at $ t=0 $ for $ \lambda = 0.1 $ where the contraction phase of universe ends up and expansion phase of universe begins. Thus, we see that the model is contracting before the bounce and begins expanding after the bounce.
 
\subsection{ Deceleration parameter} 

\qquad To explain the dynamics of universe another important parameter is deceleration parameter $ (q) $. The positive values of deceleration parameter shows the deceleration phase of the Universe whereas the negative values of $ q $ indicates the acceleration phase of the Universe. The deceleration parameter can be calculated by the relation
\begin{equation}\label{28}
q=-1-\frac{\dot{H}}{H^2}.
\end{equation}
From Eqs. (\ref{17}) and (\ref{28}), deceleration parameter $ q $ can be obtained as
\begin{equation}\label{29}
q=-\frac{\lambda  t}{6}-\frac{1}{\lambda  t}-1,
\end{equation}
which depends on time $ t $ and model parameter $ \lambda $. The plot of deceleration parameter $ q $ with respect to time $ t $ for a fixed value $ \lambda=0.1 $ is shown in Fig. 1(c). In this plot, it is clearly seen that $ q $ is transient from acceleration to deceleration before the bounce. However, it is ever accelerating model after the bounce.

\subsection{ EoS parameter}  

\qquad For the detail explanation of this bouncing model, we need to examine the other necessary conditions like behaviour of EoS parameter $ (\omega) $ and null energy condition (NEC). Therefore, it is required to find the value of energy density $ (\rho) $ and isotropic pressure $ (p) $ of the model. By solving equation (\ref {24}) and (\ref {25}) with the help of function $ F(G) $ in the form of Hubble parameter, we can obtain the values of $ \rho $ and $ p $ as
\begin{multline}\label{30}
\rho=\frac{3 (-4 \lambda ^6(\lambda  t-\frac{\lambda ^3 t^3}{3})^6+\lambda ^2 (\lambda  t-\frac{\lambda ^3 t^3}{3})^4+\lambda ^2 (1-\frac{\lambda ^2 t^2}{2})^2 (\lambda  t-\frac{\lambda ^3 t^3}{3})^2+16 \lambda ^6 (1-\frac{\lambda ^2 t^2}{2})^2 (\lambda  t-\frac{\lambda ^3 t^3}{3})^4}{\kappa  ((\lambda  t-\frac{\lambda ^3 t^3}{3})^2+(1-\frac{\lambda ^2 t^2}{2})^2)} \\
+\frac{4 \lambda ^6 (1-\frac{\lambda ^2 t^2}{2})^4 (\lambda  t-\frac{\lambda ^3 t^3}{3})^2)}{\kappa  ((\lambda  t-\frac{\lambda ^3 t^3}{3})^2+(1-\frac{\lambda ^2 t^2}{2})^2)},
\end{multline}
and
\begin{multline}\label{31}
p=\frac{12 \lambda ^6 (\lambda  t-\frac{\lambda ^3 t^3}{3})^8-3 \lambda ^2 (\lambda  t-\frac{\lambda ^3 t^3}{3})^6-2 \lambda ^2 (1-\frac{\lambda ^2 t^2}{2})^6-8 \lambda ^6 (1-\frac{\lambda ^2 t^2}{2})^8-7 \lambda ^2 (\lambda  t-\frac{\lambda ^3 t^3}{3})^2 (1-\frac{\lambda ^2 t^2}{2})^4}{{\kappa  ((\lambda  t-\frac{\lambda ^3 t^3}{3})^2+(1-\frac{\lambda ^2 t^2}{2})^2)^2}}\\ 
- \frac{8 \lambda ^2 (\lambda  t-\frac{\lambda ^3 t^3}{3})^4 (1-\frac{\lambda ^2 t^2}{2})^2-36 \lambda ^6 (\lambda  t-\frac{\lambda ^3 t^3}{3})^2 (1-\frac{\lambda ^2 t^2}{2})^6-4 \lambda ^6 (\lambda  t-\frac{\lambda ^3 t^3}{3})^4 (1-\frac{\lambda ^2 t^2}{2})^4}{\kappa  ((\lambda  t-\frac{\lambda ^3 t^3}{3})^2+(1-\frac{\lambda ^2 t^2}{2})^2)^2}\\
- \frac{28 \lambda ^6 (\lambda  t-\frac{\lambda ^3 t^3}{3})^6 (1-\frac{\lambda ^2 t^2}{2})^2}{\kappa  ((\lambda  t-\frac{\lambda ^3 t^3}{3})^2+(1-\frac{\lambda ^2 t^2}{2})^2)^2}.
\end{multline}

\begin{figure}
\begin{center}
$%
\begin{array}{c@{\hspace{.1in}}cc}
\includegraphics[width=3.0 in, height=2.2 in]{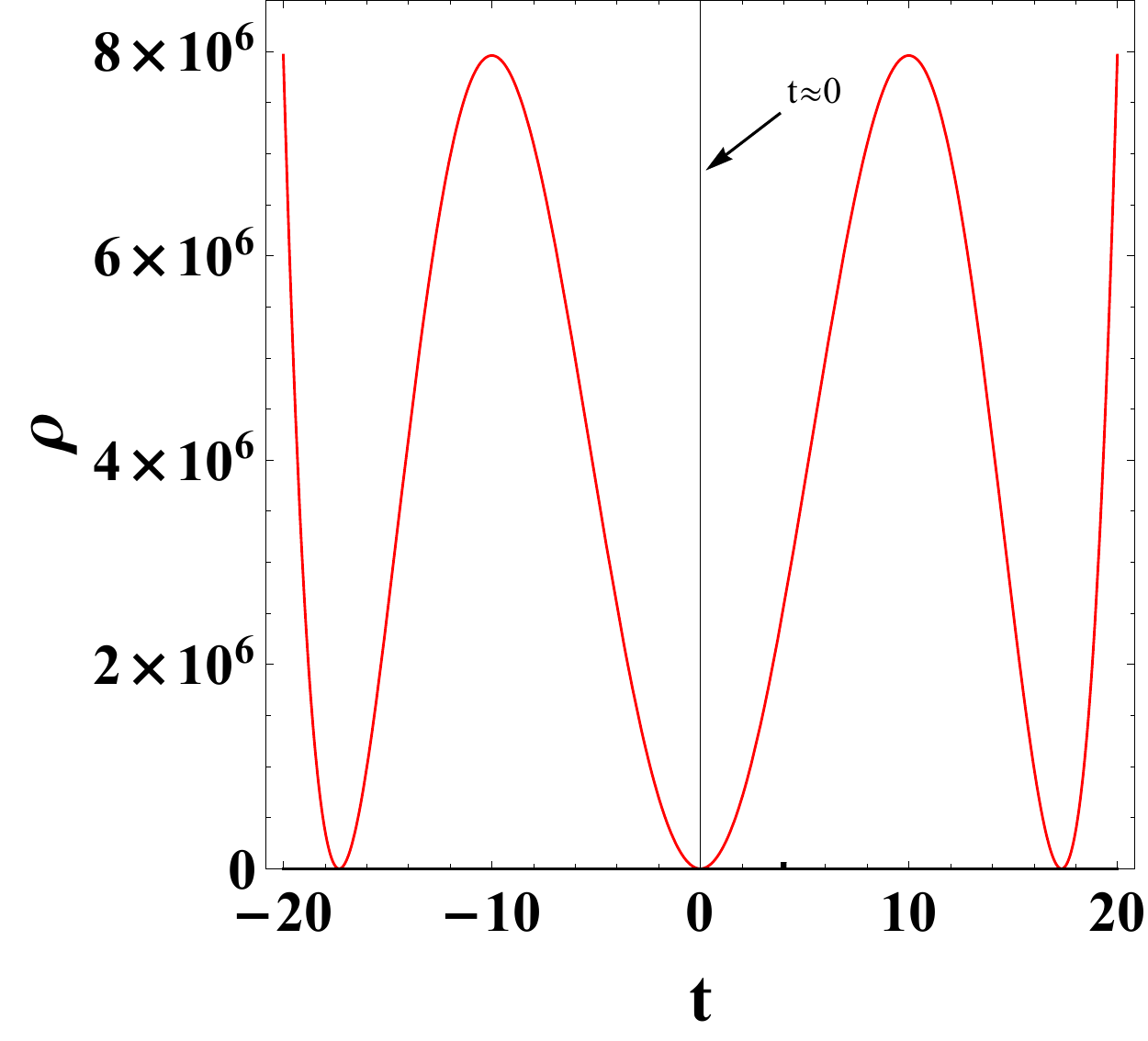} ~~~~~
\includegraphics[width=3.0 in, height=2.2 in]{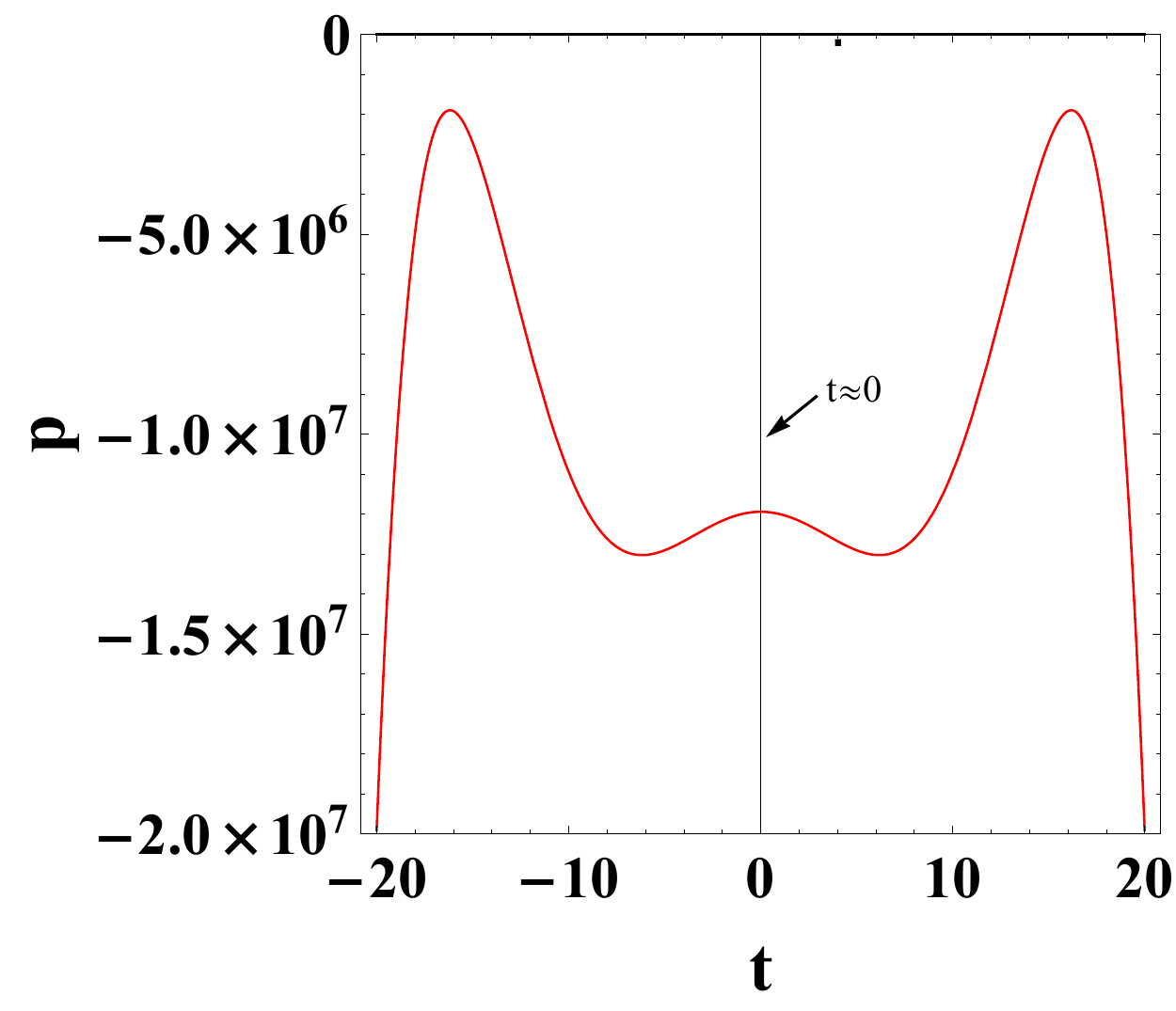} \\
\mbox (a) & \mbox (b)  \\
\includegraphics[width=3.0 in, height=2.2 in]{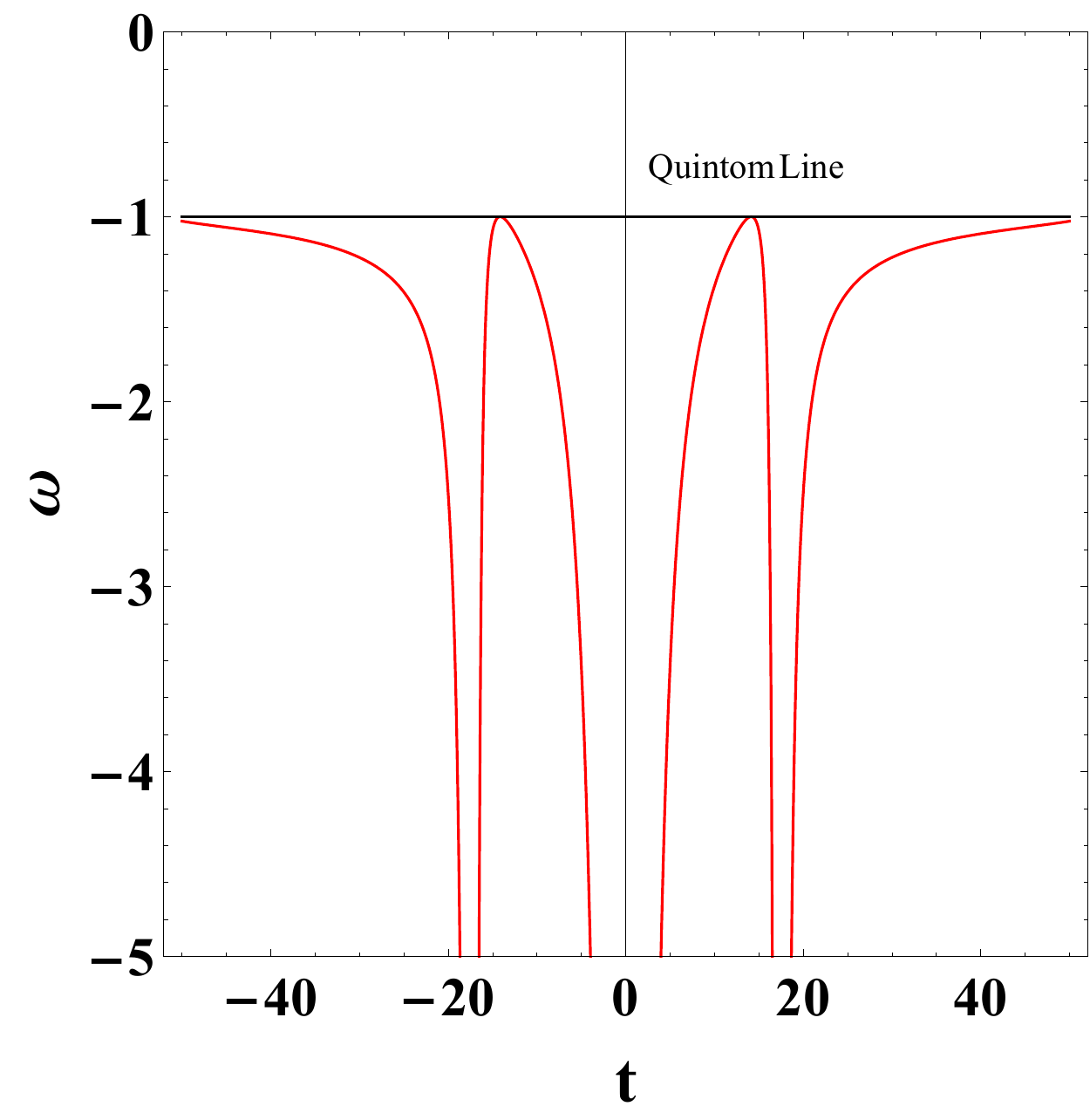} \\ 
\mbox (c)%
\end{array}%
$%
\end{center}
\caption{\scriptsize The plots of EoS parameter $ \omega $, energy density $ \rho $ and pressure $ (p) $ \textit{w.r.t.} $ t $.}
\end{figure}

The plots of pressure $ p $, energy density $ \rho $ and EoS parameter $ \omega $ for the corresponding time $ t $ for the model parameters $ \lambda=0.1 $ and $ \kappa=8\pi G=167.55 \times 10^{-11} $ can be seen in Fig. 2. In Fig. 2(a), it has been seen that energy density $ (\rho) $ approximately diminishes at the bouncing point $ t\approx0 $, it decreases before bounce and increases after the bounce. The isotropic pressure $ (p) $ is having its highly negative value for the whole range of time before and after the bounce (see Fig. 2(b)).\\ 

The EoS parameter $ (\omega) $ plays an important role in describing bouncing cosmology. In our model, the EoS parameter $ \omega $ can be calculated as

\begin{figure}
\begin{center}
$%
\begin{array}{c@{\hspace{0.1in}}c}
\includegraphics[width=3.0 in, height=2.5 in]{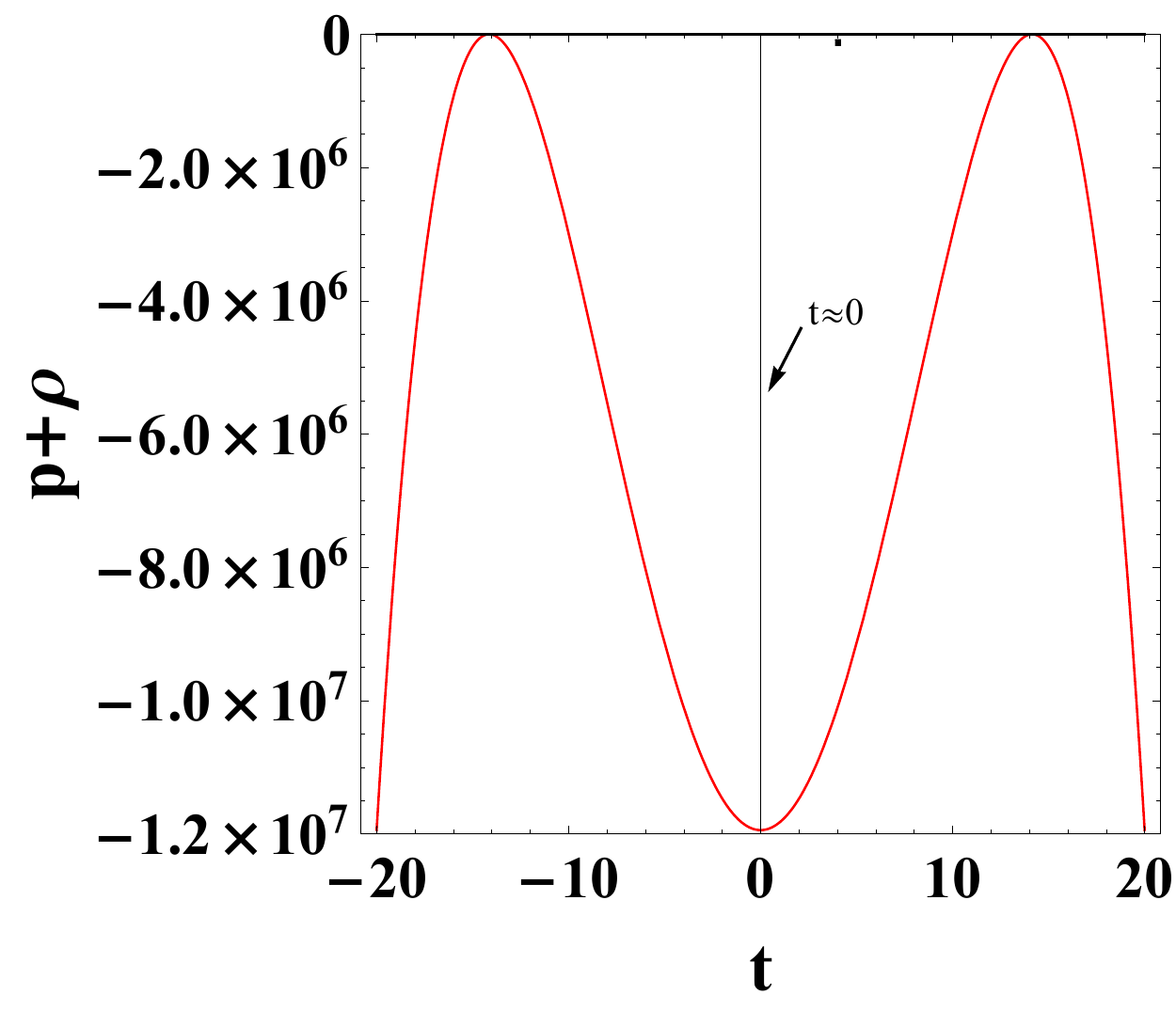} & 
\includegraphics[width=3.0 in, height=2.5 in]{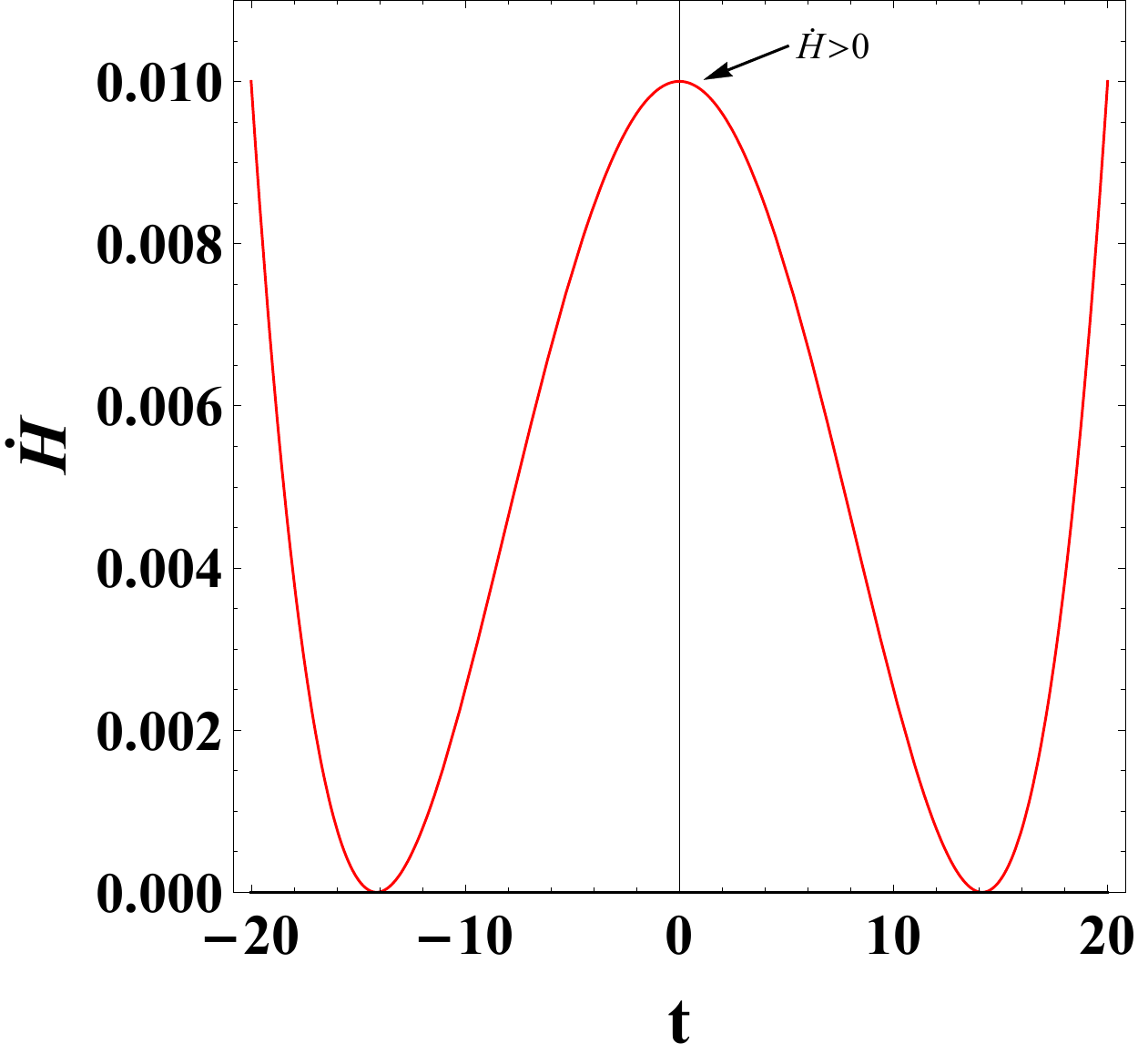} \\ 
\mbox (a) & \mbox (b)%
\end{array}%
$%
\end{center}
\caption{\scriptsize The plots of Null Energy Condition (NEC) and $ \dot{H} $ vs. $ t $.}
\end{figure}

\begin{equation}\label{32}
\omega=\frac{p}{\rho}=\frac{A+B}{C}, 
\end{equation}
where
\begin{multline}\label{33}
A=12 \lambda ^6 (\lambda  t-\frac{\lambda ^3 t^3}{3})^8-3 \lambda ^2 (\lambda  t-\frac{\lambda ^3 t^3}{3})^6-2 \lambda ^2 (1-\frac{\lambda ^2 t^2}{2})^6-8 \lambda ^6 (1-\frac{\lambda ^2 t^2}{2})^8-7 \lambda ^2 (\lambda  t-\frac{\lambda ^3 t^3}{3})^2 (1-\frac{\lambda ^2 t^2}{2})^4, ~~~~~~\\
B=-8 \lambda ^2 (\lambda  t-\frac{\lambda ^3 t^3}{3})^4 (1-\frac{\lambda ^2 t^2}{2})^2-36 \lambda ^6 (\lambda  t-\frac{\lambda ^3 t^3}{3})^2 (1-\frac{\lambda ^2 t^2}{2})^6-4 \lambda ^6 (\lambda  t-\frac{\lambda ^3 t^3}{3})^4 (1-\frac{\lambda ^2 t^2}{2})^4 \\
-28 \lambda ^6 (\lambda  t-\frac{\lambda ^3 t^3}{3})^6 (1-\frac{\lambda ^2 t^2}{2})^2,\\ 
C=3 (-4 \lambda ^6(\lambda  t-\frac{\lambda ^3 t^3}{3})^6+\lambda ^2 (\lambda  t-\frac{\lambda ^3 t^3}{3})^4+\lambda ^2 (1-\frac{\lambda ^2 t^2}{2})^2 (\lambda  t-\frac{\lambda ^3 t^3}{3})^2+16 \lambda ^6 (1-\frac{\lambda ^2 t^2}{2})^2 (\lambda  t-\frac{\lambda ^3 t^3}{3})^4 \\
+4 \lambda ^6 (1-\frac{\lambda ^2 t^2}{2})^4 (\lambda  t-\frac{\lambda ^3 t^3}{3})^2). \notag \\
\end{multline}

In this model, one can see that the EoS parameter crosses the quintom line $ \omega=-1 $ in the neighbourhood of bouncing point $ t\approx 0 $, which is a very strong criterion for a successful bouncing model \cite{Cai:2008ed}. After crossing the Quintom line $ \omega=-1 $, it has also been seen that the EoS parameter tends to the Quintom line in late times.

\subsection{ Energy Conditions}

\qquad The energy conditions (ECs) are just simple constraints on various linear combinations of the energy density and pressure. It demonstrates that energy density cannot be negative and gravity is always attractive. From the Raychaudhuri equation, the concept of energy conditions came into light. The various ECs like the null energy condition (NEC), strong energy condition (SEC), dominant energy condition (DEC), and weak energy condition (WEC) are defined as $ \rho+p \geq 0 $; $ \rho+3p \geq 0 $; $ \rho > |p| \geq 0 $; and $ \rho \geq 0 $, $ \rho+p \geq 0 $ respectively.

The successful bouncing model requires the condition $ \dot{H}=-4 \pi G \rho (1+\omega)>0 $ in the neighbourhood of the bouncing point. Thus, it is needed to see the variations of NEC and first derivative of Hubble parameter $ \dot{H}$ with respect to $ t $ in the plots of Fig. 3. In these plots, we find that NEC ($ \rho+p \geq 0 $) is violated, and $ \dot{H}>o $ in the neighbourhood of the bouncing point $ t\approx0 $, which is also a satisfactory result for a successful bouncing cosmic scenario.

\begin{figure}
\begin{center}
$%
\begin{array}{c@{\hspace{.1in}}cc}
\includegraphics[width=3.0 in, height=2.2 in]{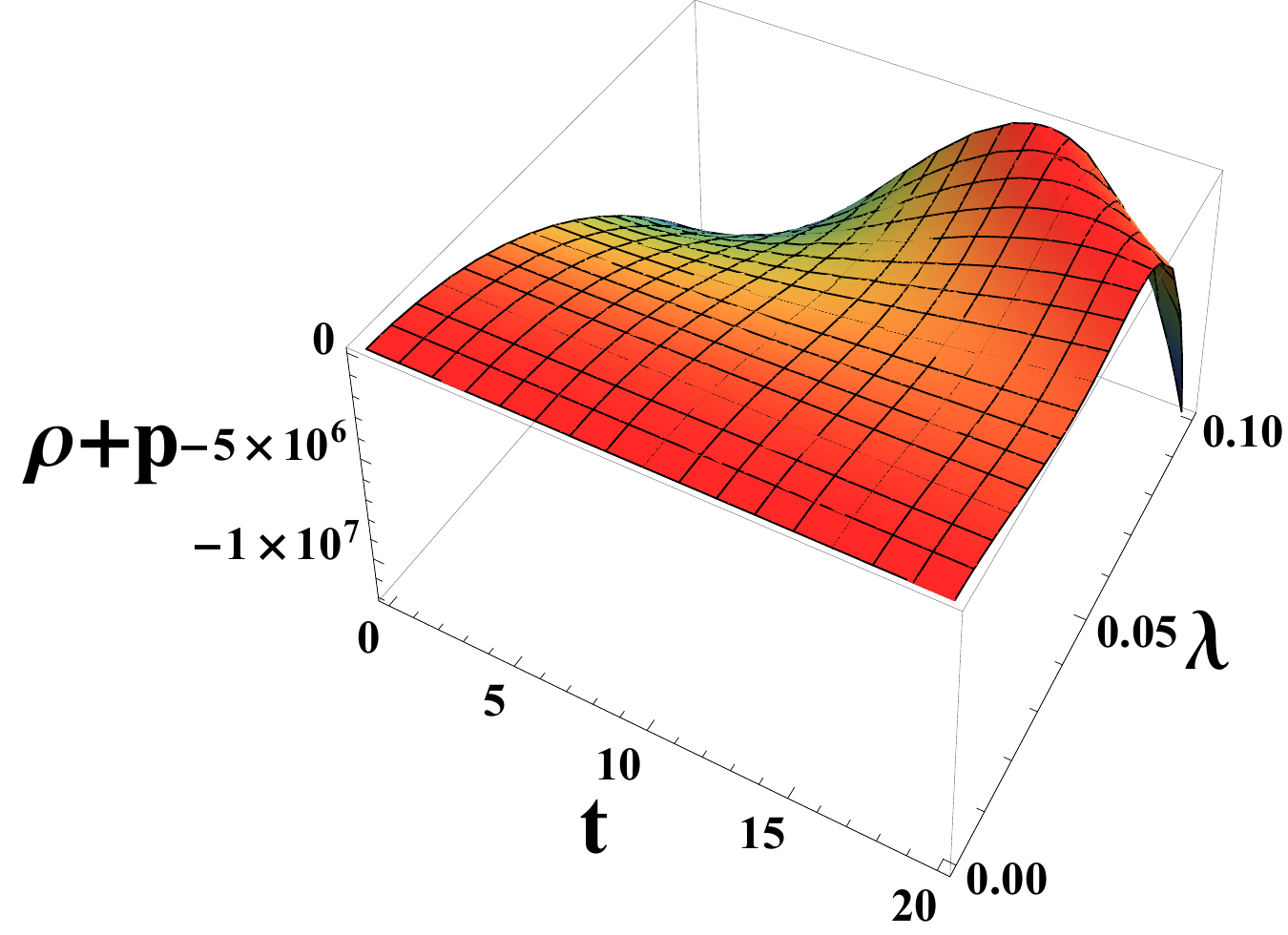} ~~~~
\includegraphics[width=3.0 in, height=2.2 in]{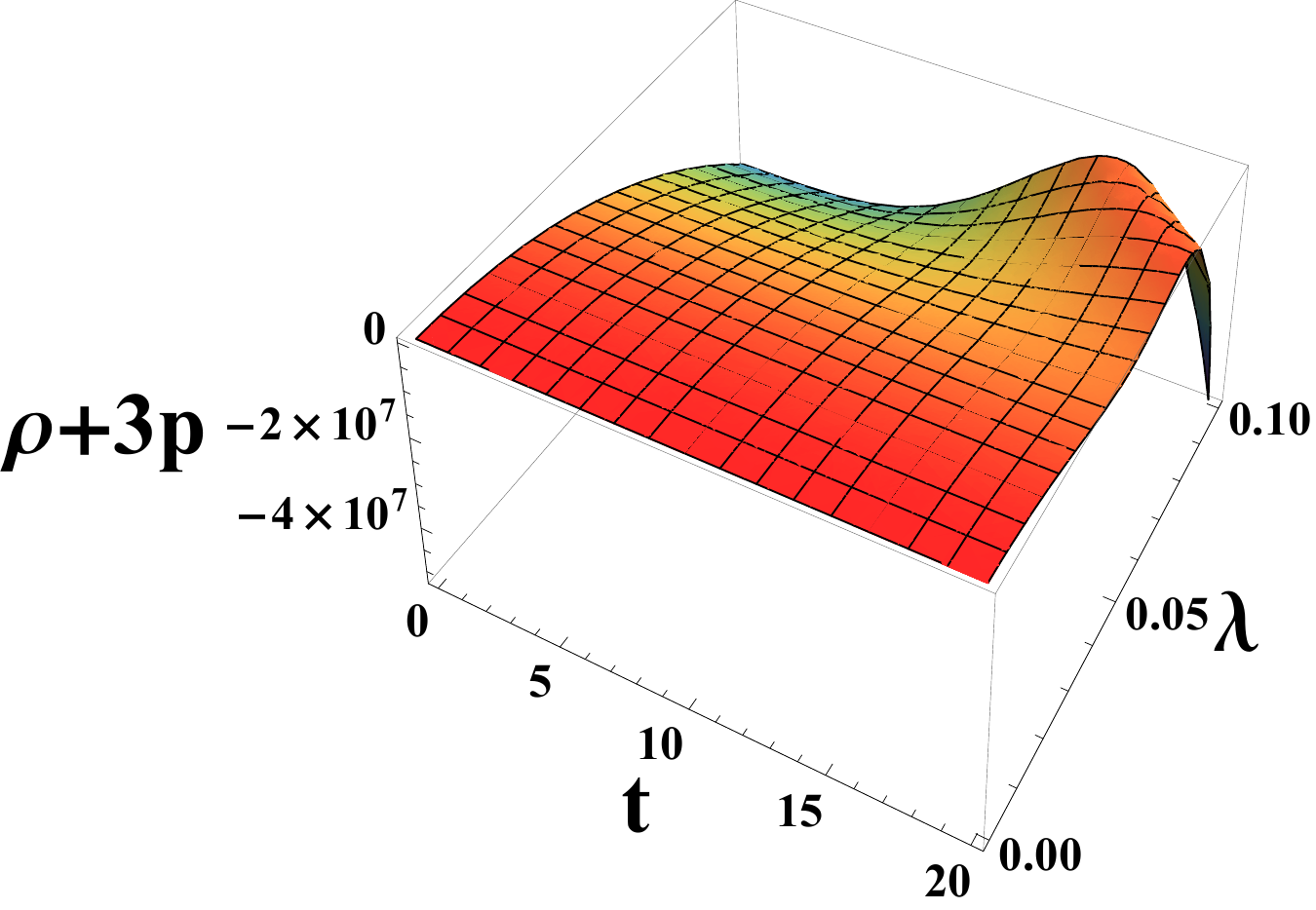} \\
\mbox (a) & \mbox (b) & \\\\
\includegraphics[width=3.0 in, height=2.2 in]{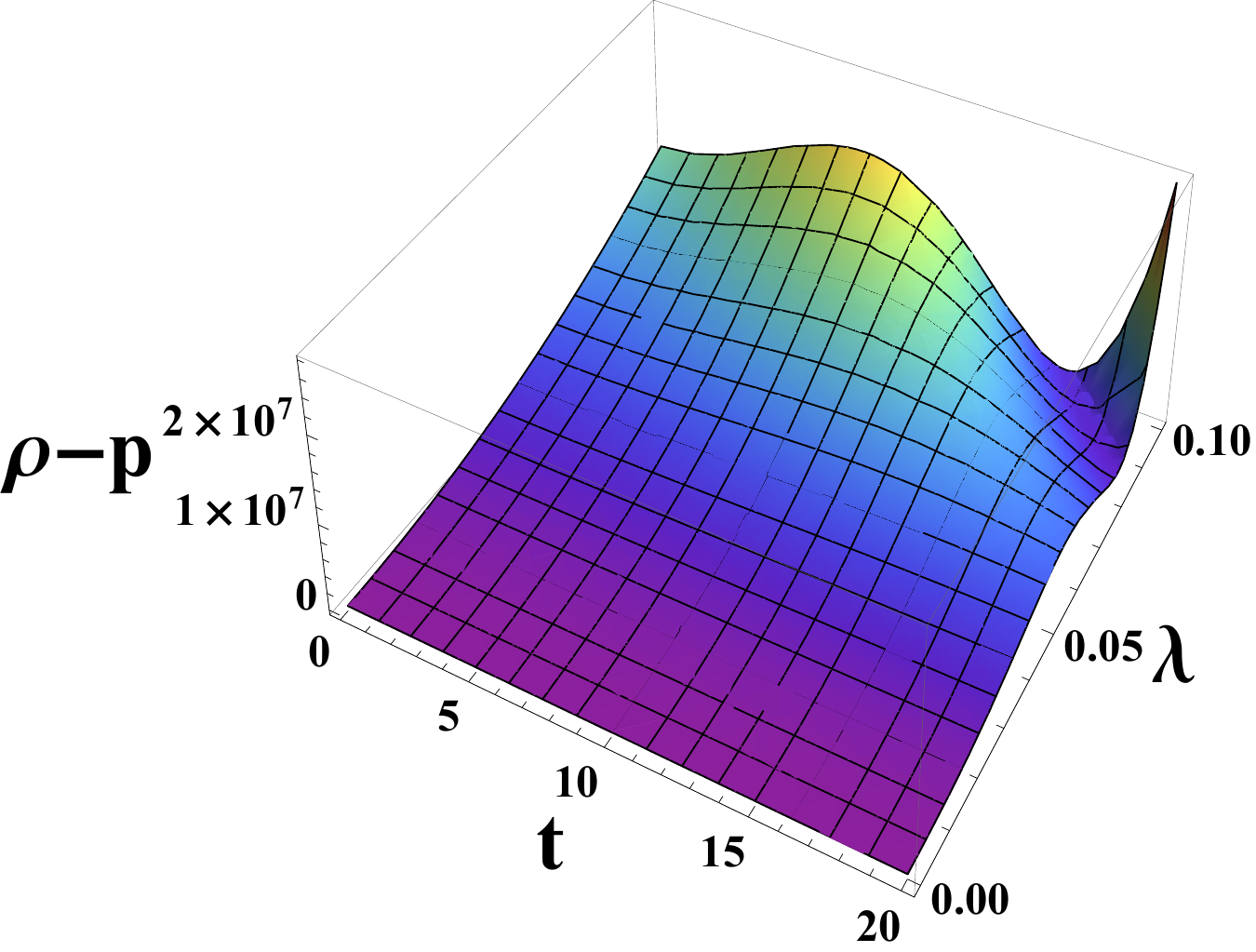} \\ 
\mbox (c)%
\end{array}%
$%
\end{center}
\caption{\scriptsize The graphical behavior of NEC, SEC and DEC.}
\end{figure}

In Fig. 4, one can observe that both NEC ($ \rho+p>0 $) and SEC ($ \rho+3p>0 $) are violating in the neighbourhood of the bouncing point $ t\approx0 $ whereas the DEC ($ \rho > |p| \geq 0 $) holds good. The violation of NEC satisfies the bouncing criteria whereas the violation of SEC indicates the existence of exotic matter in the universe. We can also analyze that DEC holds good in our model. Therefore, our model satisfies all the essential criteria for the bouncing model in $ f(R,G) $ gravity.

\section{ Conclusion}
 
\qquad In this paper, we have studied the bouncing behavior of the Universe in $ f(R, G) $ gravity. We have taken a very simple form of $ f(R, G) $ as analyzed in the literature and subsequently applied the reconstruction method by parametrizing the scale factor in the hyperbolic cosine of the cosmic time to calculate the value of $ f(G) $, and found a possible $ f(R, G) $ expression successfully. Furthermore, we have discussed the nature of the cosmological parameter which shows the bouncing scenario in our model. It is very easy to understand the bouncing cosmology with the help of the expansion rate of the universe. So, H is one of the most useful parameters which shows the contracting phase ($ H < 0 $) before the bounce, and the expanding phase ($ H > 0 $) of the Universe after the bounce at $ t\approx0 $ in our model. 

From Figs. 1, 2, 3 and 4, we observe that
\begin{itemize}
\item[(i)] The volume of the model is decreasing before the bounce and starts increasing after the bounce. The scale factor of the Universe attains to a non-zero minimum value $ a=1 $ at the bouncing point $ t\approx0 $ for $ \lambda=0.1 $ (see Fig. 1a),

\item [(ii)] The model is contracting $ H<0 $ $ H<0 $ in the interval $ -17.4<t<0 $ before the bounce and begins expanding $ H>0 $ in the interval $ 0<t<17.4 $ after the bounce, and we find $ H\approx0 $ at the bouncing point $ t\approx0 $ for $ \lambda=0.1 $ (see Fig. 1b),

\item [(iii)] In Fig. 1c, we see that the deceleration parameter $ q $ is transient from acceleration to deceleration before the bounce, and is the ever-accelerating model after the bounce,

\item [(iv)] Our obtained model is a Quintom model as the EoS parameter $\omega$ of the matter content undergoes a phase transition from $ \omega<-1 $  to $ \omega>-1 $ in the neighborhood of bouncing point at $ t\approx0 $ \cite{Cai:2007qw,Cai:2008ed} (see Fig. 2c),

\item [(v)] The null energy condition (NEC) violates, and $ \dot{H}> 0 $ in the neighbourhood of bouncing point at $ t\approx0 $ in the same interval (see Fig. 3).

\item [(vi)] The violation of NEC satisfies the bouncing criteria whereas the violation of SEC indicates the existence of exotic matter in the universe (see Fig. 4).
\end{itemize}

Thus, we find that the EoS parameter crosses quintom line ($ \omega=-1 $), $ \dot{H} > 0  $, and violating the null energy condition ($ \rho+p>0 $) in the neighbourhood of bouncing point. In our model, we have also observed that a universe avoids the Big Bang singularity in the presence of the Quintom matter. Thus, we conclude that our presented model is a bouncing model in $ f(R, G) $ gravity which is consistent with the models discussed by several authors \cite {Battefeld:2014uga,Cai:2007qw,Ijjas:2016tpn} etc. 

\vskip0.2in 
\textbf{\noindent Acknowledgements} JKS and Shaily express their thanks to Prof. S. G. Ghosh, CTP, Jamia Millia Islamia, New Delhi, India and Manuel Malheiro, Physics Department of ITA, Brasil for fruitful discussions and suggestions. The work of author KB was supported in part by the JSPS KAKENHI Grant Number JP21K03547. Authors also express their thanks to the referee for his valuable comments and suggestions. 

\end{document}